\documentstyle[prb,twocolumn,aps]{revtex}
\begin{document}
\twocolumn[\hsize\textwidth\columnwidth\hsize\csname @twocolumnfalse\endcsname

\title{Coulomb Driven New Bound States at the Integer Quantum Hall 
States in GaAs/Al$_{0.3}$Ga$_{0.7}$As Single Heterojunctions} 
\author{Yongmin\, Kim$^*$, F. M.\, Munteanu$^{*,\dag}$, C. H.\, 
Perry$^{*,\dag}$, X.\, Lee$^{\ddag}$, H. W.\, Jiang$^{\ddag}$, J. A.\, 
Simmons$^{\S}$, and Kyu-Seok\, Lee$^{\P}$} 

\address{$^*$National High Magetic Field Laboratory-Los Alamos 
National Laboratory, Los Alamos, NM 87545\\
$^{\dag}$Department of Physics, Northeastern University, Boston, MA 
02115\\ 
$^{\ddag}$Department of Physics and Astronomy, University of 
California, Los Angeles, CA 90095\\
$^{\S}$Sandia National Laboratory, Albuquerque, NM 87185\\
$^{\P}$Electronics and Telecommunications Research Institute, Yuseong, 
Korea}
\date{\today} \maketitle
\begin{abstract}
Coulomb driven, magneto-optically induced electron and hole bound 
states from a series of heavily doped GaAs/Al$_{0.3}$Ga$_{0.7}$As 
single heterojunctions (SHJ) are revealed in high magnetic fields.  At 
low magnetic fields ($\nu>$2), the photoluminescence spectra display 
Shubnikov de-Haas type oscillations associated with the empty second 
subband transition.  In the regime of the Landau filling factor 
$\nu<$1 and 1$<\nu<2$, we found strong bound states due to Mott type 
localizations.  Since a SHJ has an open valence band structure, these 
bound states are a unique property of the dynamic movement of the 
valence holes in strong magnetic fields.
\end{abstract}
%\pacs{Valid PACS appear here.
%{\tt$\backslash$\string pacs\{\}} should always be input,
%even if empty.}

]\narrowtext

In the last several years, many optical studies focused on the regime 
of the integer and fractional quantum Hall states of semiconductor 
quantum wells (QWs), where electrons and holes have confined energy 
levels.\cite{perry,heiman,turberfield,chen,kukushkin,hawrylak,gravier,kim} 
Since a SHJ has only one interface, it is easier to fabricate high 
quality devices with ultrahigh mobilities.  In a SHJ, the conduction 
band electrons are confined in a wedge-shaped quantum well near the 
interface, whereas the photocreated valence holes are not confined and 
tend to move to the GaAs flat band region.  This has often been 
considered to be a disadvantage in optical experiments since the 
dynamic movement of valence holes due to the open structure makes it 
difficult to judge their location.  To avoid this problem, intentional 
acceptor doping techniques have been employed to study optical 
transitions from SHJs.\cite{kukushkin,hawrylak} For example, 
magnetophotoluminescence (MPL) experiments have been carried out on 
acceptor doped SHJs to investigate transitions associated with the 
integer quantum Hall effect (IQHE) and fractional quantum Hall effect 
(FQHE).\cite{kukushkin,hawrylak}

In this Letter, we report the observation of strong discontinuous 
transitions at the $\nu$=2 and $\nu$=1 integer quantum Hall states and 
we believe these new transitions are the consequence of the formation 
of Coulomb-driven strong bound states.  In a heavily doped single 
heterojunction, there are two factors that enhance the second subband 
(E1) exciton transition.  The first is due to the close proximity 
between the Fermi energy and the E1 subband;\cite{chen,kim} the second 
is due to the spatially indirect nature of conduction and valence band 
structure, the wavefunction overlap between the E1 subband and the 
valence hole is much larger than that of the first subband (E0) and 
the valence hole.\cite{chen} As the magnetic field changes, the 2DEG 
modifies the valence hole self-energy.\cite{kim,uenoyama} This in turn 
causes the E1 exciton transition to display strong oscillatory 
behavior in its transition energy and peak intensity at magnetic 
fields smaller than the $\nu$=2 integer quantum Hall 
state.\cite{chen,kim} Near $\nu$=2, however, the E1 exciton transition 
loses its intensity and a new feature emerges at a lower energy.  This 
transition rapidly increases its intensity for 2$>\nu>$1 but then 
diminishes in strength and disappears as the magnetic field approaches 
the $\nu$=1 quantum Hall state.  In its place, another new peak 
appears at lower energy which swiftly grows in intensity for $\nu<$1.  
These two red-shifted transitions are attributed to the formation of 
new bound states due to electron and hole localization and can still 
be observed to temperatures as high as 40K. In addition, we found that 
the strong bound states at $\nu$=2 and $\nu$=1 were not observed for 
low electron density samples (sample 1 and 2).  These samples have a 
relatively large separation between the Fermi energy and the E1 
subband and only the E0-hole free carrier transitions were observed.

A series of samples grown by different growth techniques was used for 
this study.  One set was grown using a metal-oxide chemical vapor 
deposition (MOCVD) technique\cite{chui} and the other group was 
fabricated using a molecular beam epitaxy (MBE) technique.  Table 1 
shows the various parameters for six samples.  The high magnetic 
fields were generated using a recently commissioned 60T 
quasi-continuous (QC) magnet, which has 2-second field duration.  A 
pumped $^{3}$He cryostat were used to achieve temperatures of 0.4-70K. 
For MPL experiments, a 630nm low power diode laser ($<$1.5mW/cm$^{2}$ 
to the samples) was used as the excitation source and a single optical 
fiber (600mm diameter; 0.16 numerical aperture) provided both the 
input excitation light onto sample and the output PL signal to the 
spectrometer.\cite{perry2} The spectroscopic system consisted of a 300 
mm focal length f/4 spectrometer and a charge coupled device (CCD) 
detector, which has a fast refresh rate (476Hz) and high quantum 
efficiency (90\% at 500nm).  This fast detection system allowed us to 
collect approximately 500 PL spectra/second during the duration of the 
magnet field pulse.

The $\sigma$+ polarization MPL spectra for sample 3 as a function of 
magnetic field taken in the QC magnet is displayed in Fig.  1.  The 
corresponding MPL intensities for both the $\sigma$+ and $\sigma$- 
polarizations are shown in Fig.  2.  Two dramatic intensity changes 
with high magnetic fields near the $\nu$=2 and $\nu$=1 integer quantum 
Hall states were observed for these heavily doped samples.  Our 
self-consistent calculations of the conduction band energies (Fig.  3) 
indicate that the Fermi energy lies close proximity to the E1 subband 
for heavily doped samples and the discontinuous MPL transitions are 
associated with the Fermi edge enhanced E1 exciton 
transitions.\cite{chen} The energy shifts of the peaks are displayed in 
Fig.  4a as a function of magnetic field B. As the magnetic field 
increases further, the transition (P2) that occurred at $\nu$=2 
disappears near $\nu$=1 and another transition (P1) emerges at the 
$\nu$ =1 quantum Hall state.  Fig.  4b shows oscillatory behaviors of 
the MPL transitions at low magnetic fields.

As indicated in Fig.  3 for heavily doped SHJs, the Fermi energy lies 
close to the E1 subband ($<$1meV) which induces E1 exciton 
transition.\cite{chen} The emergence of the P2 transition at the 
$\nu$=2 quantum Hall state can be explained as follows: At low fields 
($\nu>$2), valence holes are unbound and are free to move to the GaAs 
flat band region due to the open structure of the valence band (see 
Fig.  2 inset).  As the magnetic field varies, the Fermi energy sweeps 
continuously from localized states to extended states, which changes 
the dielectric screening of the valence holes.  When this happens, 
there are two factors to be considered in an optical transition.  One 
is the hole self-energy and the other is the vertex correction arising 
from the exciton effect.  The hole self-energy gives rise to a blue 
shift,\cite{kim,uenoyama} whereas the vertex correction term gives a 
red-shift due to the exciton binding energy.  Since the hole 
self-energy correction is bigger than the exciton 
effect\cite{uenoyama} the free carrier transition shows a blue shift 
at even Landau filling factors.  For the E1 exciton transition 
measured in these heavily doped samples, the valence hole self-energy 
is already modified by the E0 free carriers and hence it shows a blue 
shift at even filling factors for $\nu>$2 (see Fig.  4b).  Unlike 
transitions at even filling factors for $\nu>$2, the P2 transition at 
$\nu$=2 shows a large red-shift.  This means that the vertex 
correction at $\nu$=2 in a heavily doped SHJ is much larger than the 
other effects which give a blueÐshift in the transition energy.  At 
the $\nu$=2 quantum Hall state, electrons are strongly localized and 
the screening within the 2DEG becomes negligible.  In the absence of 
screening effects, holes migrate back towards the interface because of 
the strong Coulomb attraction between the electrons and holes.  As a 
result, holes are localized near the interface and form a new bound 
state (P2) at the $\nu$=2 quantum Hall state.  This Coulomb-driven 
hole localization induces a large vertex correction that gives rise to 
a giant red-shift in the transition energy.  For sample 3, the P2 
transition has a binding energy of about 8.2meV which decreases to 
~5meV at $\nu$=1 where it disappears.  The amount of the red-shift 
would correspond to binding energy of the new bound state.  The 
red-shift in this field regime is analogous to a Mott type 
transition,\cite{yoon,gekhtman} since it is similar to the metal to 
insulator transition in the hydrogen system originally suggested by 
Mott,\cite{mott} for which the transition is induced when the screening 
length exceeds a critical value caused by the expanding lattice 
constant.  The relative binding energy ({\it$\Delta$$E2$}) of the new 
bound state (P2) increases with increasing 2DEG density.  This is 
consistent with the reduction in the screening at the $\nu$=2 QHS. 
When the screening is turned off, the higher density sample has the 
stronger Coulomb attraction between electrons and holes.  Hence, a 
higher 2DEG density sample has a larger binding energy than a lower 
electron density sample.

Near $\nu$=1, the P2 transition disappears and a new peak 
designated as P1 emerges on the lower energy side of the P2 
transition.  Like the P2 transition, the P1 transition also rapidly 
increases in intensity with increasing magnetic field as seen in Fig.  
2.  Near $\nu$=1, the screening strength within the 2DEG in the well 
is greatly reduced and once again the localization of the valence hole 
induces the discontinuous transitions and intensity changes at 
$\nu<$1. As indicated in Table 1, the amount of the $\Delta$$E1$ and 
the $\Delta$$E2$ measured at $\nu$=1 and $\nu$=2, respectively, are 
almost the same for a given 2DEG density.

There are numerous magneto-optical studies near the $\nu$=1 filling 
state where a red-shift in transition energy has been observed as a
`shake-up' process at the Fermi 
energy.\cite{gravier,nicholas,cooper,osborne} However, our circularly 
polarized MPL measurements show completely different results from 
others.\cite{gravier,osborne} In our experiments, we found that both 
the P1 and P2 transitions were strongly left circularly polarized 
($\sigma$-) as seen in Fig.  2.  The intensity of the P2 $\sigma$- 
transition is about three times that of the $\sigma$+ transition, 
whereas P1 has about a 5:1 ratio for $\sigma$-/$\sigma$+.  The E1 
transition, on the other hand, shows no appreciable intensity 
differences between the two spin polarizations.  This means that for an 
unbound exciton state (E1), both spins are almost equally populated, 
whereas for the strongly bound exciton states (P1 and P2), they are 
strongly polarized to the excitonic ground state ($\sigma$-).  Though 
not shown here, the temperature dependence of the MPL experiments show 
that the P2 transition disappears at ~40K, whereas the P1 transition 
disappears at ~10K. This is due to the fact that the thermal broadening of 
the 2DEG density of states closes the Zeeman gap of the Landau levels 
which does not contribut the reduction of the screening at the odd 
integer filling states at ~10K.

It has been suggested\cite{kivelson,jiang,wang} that within the 
quantum Hall phase diagram, the mobility of the sample strongly 
affects the quantum Hall liquid (QHL) to quantum Hall insulator (QHI) 
phase transition, as the induced electron localization can take place 
at different quantum Hall states.  For example, for a highly 
disordered structure the phase transition occurs near $\nu$=2, but for 
a moderately disordered one the phase transition is near $\nu$=1.  As 
the samples used in this study have high mobilities 
($>$10$^6$cm$^{2}$/Vs), we may expect to see other bound states near 
$\nu$=1/3 fractional quantum Hall state caused by the quantum Hall 
phases transition.  This would manifest as another minimum in the 
intensity for sample 3 at about 69T for $\nu$=1/3.  Unfortunately, 
this is just beyond the current high field limit of 60T for the QC 
magnet at NHMFL-LANL but there is some evidence that this may be 
about to take place as the intensity of the P1 transition shows a 
rapid decrease between 50 and 60T (see Fig.  2).

We have presented MPL studies on a series of high mobility 
SHJs in high magnetic fields to 60T using a QC magnet at NHMFL-LANL. 
At low magnetic fields ($\nu>$2), the photoluminescence spectra 
display Shubnikov de-Haas type oscillations associated with the empty 
second subband transition.  In the high field regime, we observe the 
formation of Coulomb driven, magneto-optically induced electron and 
hole bound states near Landau filling factors $\nu$=2 and $\nu$=1 
(with some evidence that there may be another at $\nu$=1/3).  A 
discrete phase transformation from a dynamic hole to a bound hole 
state due to a Mott-type transition is thought to be responsible for 
the large red-shift that occurs near the $\nu$=2 and $\nu$=1 Landau 
filling states.  Both bound state transitions are strongly spin 
polarized ($\sigma$-) states.  The appearance of these bound states 
appears to be a unique property of heavily doped SHJs and the 
associated dynamic movement of the holes in strong magnetic fields due 
to their open valence band structure.

Authors gratefully acknowledge the engineers and technicians at 
NHMFL-LANL for their efforts on operating 60T QC magnet.  Work at 
NHMFL-LANL is supported by NSF Cooperative Agreement DMR 9527035 and 
US DOE. Work at Sandia National Laboratories and UCLA is supported by 
DOE under Contract DE-AC04-94AL85000 and by NSF Cooperative Agreement 
DMR 9705439, respectively.

\begin{figure}
\caption{Approximately 1000 MPL spectra taken in a 2-second magnetic 
field sweep of the 60T QC magnet for sample 3 at T=1.5K for $\sigma$+ 
polarization.  The peaks labeled P1, P2 and E1 are described in the 
text.}
\label{Fig1}
\end {figure}

\begin{figure}
\caption{A plot of the PL transition intensity vs magnetic field for 
sample 3.  The E1 transition intensity rapidly diminishes for 
$\nu<$2.  At $\nu$=2 and $\nu$=1, two new transitions assigned as P2 and P1, 
respectively, emerge on the lower energy side of the E1 transition.  
Compared with the E1 transition, which is unpolarized, the P2 and P1 
transitions are strongly polarized ($\sigma$-/LCP).  The inset depicts a SHJ 
structure and the inferred movement of the valence hole at $\nu<$2.}
\label{Fig2}
\end{figure}

\begin{figure}
\caption{The results of self-consistent calculations for SHJ subband 
energy levels with respect to the spacer thickness under laser 
illumination. The vertical arrows indicate the samples used for this 
study. The right axis is the energy difference ($\Delta$) between the 
Fermi energy and the E1 subbnad.}
\label{Fig3}
\end {figure}

\begin{figure}
\caption{(a) A plot of the transition energies vs magnetic field for E1, 
P1 and P2 for sample 3.  Note the large discontinuous transitions at 
$\nu$=2 and $\nu$=1. (b) Expanded view of E1 vs.  magnetic field, showing the 
Shubnikov de-Haas type oscillations in the transition energy and 
intensity. The blue-shifts in the transition energy at even integer 
filling states are due to the variation of the hole self-energy. The 
intensity oscillations are due to the interaction between the Fermi 
energy and E1 subband (see text).}
\label{Fig4}
\end {figure}

\begin{table}
\caption{Sample parameters. The electron density ($n_{2D}$) under 
laser illumination was determined by transport experiments. The 
intensity oscillations in Fig. 3 are in phase with measured Shubnikov 
de-Haas oscillations. $\Delta$$E1$ and $\Delta$$E2$ are the values 
at $\nu$=1 and $\nu$=2, respectively, and these are almost the same 
for a given $n_{2D}$.}
\begin{tabular}{lccccc}
 &$n_{2D}$(dark)&$n_{2D}$(illu)&$\Delta$$E2$&$\Delta$$E1$&growth \\ 
 &($10^{11}/cm^{2}$)&($10^{11}/cm^{2}$)&(meV)&(meV)&technique\\ \tableline
Sample 1 &2.1&3.0 &- &- &MOCVD\\
Sample 2 &2.3&3.5 &- &- &MOCVD\\
Sample 3 &3.6&5.8 &8.4 &8.0 &MOCVD\\
Sample 4 &4.5&7.2 &9.4 &9.4 &MOCVD\\
Sample 5 &3.1&6.2 &8.6 &8.5 &MBE\\
Sample 6 &-&14.4 &12.0 &- &MBE\\
\end{tabular}
\end{table}


\begin{references}
	
\bibitem{perry} C. H. Perry, J. M. Worlock, M. C. Smith, and A. 
Petrou, in {\it High Magnetic Fields in Semiconductor Physics}, edited 
by G. Landwehr (Springer-Verlag, Berlin, 1987) 
\bibitem{heiman} D. Heiman, B. B. Goldberg, A. Pinczuk, C. W. Tu, 
A. C. Gossard, and J. H. English, Phys.  Rev.  Lett. {\bf 61}, 605 (1988)
\bibitem{turberfield} A. J. Turberfield, S. R. Haynes, P. A. Wright, 
R. A. Ford, R. G. Clark, J. F. Ryan, J. J. Harris, and C. T. Foxon, 
Phys.  Rev.  Lett.  {\bf 65}, 637 (1990) 
\bibitem{chen} W. Chen, M. Fritze, A. V. Nurmikko, D. Ackly, C. 
Colvard, and H. Lee, Phys.  Rev.  Lett.  {\bf 64}, 2434 (1990) 
\bibitem{kukushkin} I.V. Kukushkin, R. J. Haug, K. von Klitzing, and 
K. Ploog, Phys.  Rev.  Lett.  {\bf 72}, 736 (1994)
\bibitem{hawrylak} P. Hawrylak, N. Pulsford and K. Ploog, Phys.  
Rev.  B {\bf 46}, 15193 (1992) 
\bibitem{gravier} L. Gravier, M. Potemski, P. Hawrylak, and B. 
Etienne, Phys.  Rev.  Lett.  {\bf 80}, 3344 (1998) 
\bibitem{kim} Y. Kim, C.H. Perry, K. -S. Lee, and D. G. Rickel, Phys. 
Rev. B {\bf 59}, 1641 (1999) 
\bibitem{uenoyama}T. Uenoyama and L. J. Sham, Phys.  Rev.  B {\bf 39}, 
11044 (1989) 
\bibitem{chui} H. C. Chui, B. E. Hammons, N. E. Harff, J. A. Simmons, 
and M. E. Sherwin, Appl.  Phys.  Lett.  {\bf 68} (2), 208 (1996) 
\bibitem{perry2}C.H. Perry, Y. Kim, and D. G. Rickel, Physica B {\bf 246-247}, 
182 (1998)
\bibitem{yoon} H. W. Yoon, M. D. Sturge, and L. N. Pfeiffer, Solid State Comm.  
Vol.  {\bf 104}, 287 (1997)
\bibitem{gekhtman} D. Gekhtman, E. Cohen, A. Ron, and L. N. 
Pfeiffer, Phys.  Rev.  B {\bf 54}, 10 320 (1996) 
\bibitem{mott} N. F. Mott, Proc.  Phys.  Soc.  (London) {\bf 62}, 1949, 416 
\bibitem{nicholas} R. J. Nicholas, D. Kinder, A. N. Priest, C. C. Chang, H. H. 
Cheng, J. J. Harris, and C. T. Foxon, Physica B {\bf 249-251}, 553, (1998) 
\bibitem{cooper}N. R. Cooper and D. B. Chklovskii, Phys.  Rev.  B {\bf 55}, 
2436 (1997) 
\bibitem{osborne}J. L. Osborne, A. J. Shields, M. Y. Simmons, N. R. 
Cooper, D. A. Ritchie, M. Pepper,  Physica B {\bf 249-251}, 538, (1998) 
\bibitem{kivelson} S. Kivelson, D. -H. Lee, and S. -C. Zhang, Phys.  Rev.  B 
{\bf 46}, 2223 (1992) 
\bibitem{jiang} 14.  H. W. Jiang, C. E. Johnson, K. L. Wang, and S. T. Hannahs, 
Phys.  Rev.  Lett.  {\bf 71}, 1439 (1993); 
\bibitem{wang} Y. J. Wang, B. D. McComb, R. Meisels, F. Kuchlar, and W. Schaff, Phys.  
Rev.  Lett.  {\bf 75}, 906 (1995)
\end{references}
\end{document}